\documentstyle[11pt,amssymb,epsfig,psfig]{article}

\begin{document}

\title{{\bf Casimir effect in Domain Wall formation}}

\author{
 M. R. Setare  \footnote{E-mail: rezakord@yahoo.com} \\
  { Institute for Theoretical Physics and Mathematics, Tehran,
Iran} \\ {Department of Science, Physics Group, Kurdistan
University, Sanandeg, Iran}\\{Department of Physics, Sharif
University of Technology, Tehran, Iran }}

\maketitle

\begin{abstract}
The Casimir forces on two parallel plates in conformally flat de
Sitter background due to conformally coupled massless scalar field
satisfying mixed boundary conditions on the plates is
investigated. In the general case of mixed boundary conditions
formulae are derived for the vacuum expectation values of the
energy-momentum tensor and vacuum forces acting on boundaries.
 Different cosmological constants are
assumed for the space between and outside of the plates to have
general results applicable to the case of domain wall formations
in the early universe.
\end{abstract}

\newpage

\section{Introduction}

The Casimir effect is one of the most interesting manifestations
  of nontrivial properties of the vacuum state in quantum field
  theory [1,2]. Since its first prediction by
  Casimir in 1948\cite{Casimir} this effect has been investigated for
  different fields having different boundary geometries[3-5]. The
  Casimir effect can be viewed as the polarization of
  vacuum by boundary conditions or geometry. Therefore, vacuum
  polarization induced by a gravitational field is also considered as
  Casimir effect.\\
  In the context of hot big bang cosmology, the unified theories
  of the fundamental interactions predict that the universe passes
  through a sequence of phase transitions. These phase transitions
  can give rise to domain structures determined by the topology of
  the manifold $M$ of degenerate vacuua \cite{{zel},{kib},{viel}}. If
  $M$ is disconnected, i.e. if $\pi(M)$ is nontrivial, then one
  can pass from one ordered phase to the other only by going
  through a domain wall. If $M$ has two connected components, e.g.
  if there is only a discrete reflection symmetry with
  $\pi_{0}(M)=Z_{2}$, then there will be just two ordered phase
  separated by a domain wall. In the domain wall formation models, in the
   early universe, the space-time changes from de Sitter to the
   geometry induced by the presence of a domain wall. In \cite{And}
   the effects of particle production and vacuum polarization attendant
   to the domain wall formation have been studied. Casimir stress for parallel
   plates in the background of static domain wall in four and two dimensions
  is calculated in \cite{{set1},{set2}}. Spherical bubbles immersed in different de Sitter
  spaces in- and out-side is calculated in \cite{{set3},{set4}}.  \\
  Our aim is to calculate the Casimir stress on two parallel plates with
  constant comoving distance having different vacuums between and outside, i.e. with
   false/true vacuum  between/outside. Our model may be used to study the effect of
   the Casimir force on the dynamics of the domain wall formation appearing in
  the simplest Goldston model. In this model potential of the scalar field
  has two equal minima corresponding to degenerate vacuua. Therefore, scalar field
  maps points at spatial infinity in physical space nontrivially into the
  vacuum manifold \cite{vil1}. Domain walls occur at the boundary between
  these regions of space. One may assume that the outer regions of parallel plates are
  in $\Lambda_{out}$ vacuum corresponding to degenerate vacuua in domain
  wall configuration. Previously this method has been used in
  \cite{set5} to drive the vacuum stress on parallel plates for scalar field with Dirichlet
  boundary condition in de Sitter space. For Neumann or more general mixed
boundary conditions we need to have the Casimir energy-momentum
tensor for the flat spacetime counterpart in the case of the Robin
boundary conditions with coefficients related to the metric
components of the brane-world geometry and the boundary mass
terms. The Casimir effect for the general Robin boundary
conditions on background of the Minkowski spacetime was
investigated in Ref. \cite{sah} for flat boundaries. Here we use
the results of Ref. \cite{sah} to generate vacuum energy--momentum
tensor for the plane symmetric conformally flat backgrounds. Also
this method has been used in \cite{set6} to derive the vacuum
characteristics of the Casimir configuration on background of
conformally flat brane-world geometries for massless scalar field
with Robin boundary condition on plates.\\
   In section two we calculate the stress on
  two parallel plates with Robin boundary conditions. The case of
  different de Sitter vacuua between and outside of the plates, is
  considered in section three. The last section conclude and
  summarize the results.

\section{Vacuum expectation values for the energy-momentum tensor}
We will consider a conformally coupled massless scalar field $%
\varphi (x)$ satisfying the equation
\begin{equation}
\left( \nabla _{\mu }\nabla ^{\mu }+\frac{1}{6} R\right) \varphi
(x)=0, \label{fieldeq}
\end{equation}
on background of a de Sitter space-time. In Eq. (\ref{fieldeq})
$\nabla _{\mu }$ is the operator of the covariant derivative, and
$R$ is the Ricci scalar for the de Sitter space.
\begin{equation}
R=\frac{12}{\alpha^{2}},  \label{Riccisc}
\end{equation}
To make the maximum use of the flat space calculation we use the
de Sitter metric in the conformally flat form:
\begin{equation}
ds^{2}=\frac{\alpha^{2}}{\eta^{2}}
[d\eta^{2}-\sum_{\imath=1}^{3}(dx^{\imath})^{2}],
\end{equation}
where $\eta$ is the conformal time:
\begin{equation}
\infty < \eta < 0.
\end{equation}
The relation between parameter $\alpha$ and cosmological constant
$\Lambda$ is given by
\begin{equation}
\alpha^{2}=\frac{3}{\Lambda}.
\end{equation}
We will assume that the field satisfies the mixed boundary
condition
\begin{equation}
\left( a_{j}+b_{j}n^{\mu }\nabla _{\mu }\right) \varphi
(x)=0,\quad z=z_{j},\quad j=1,2  \label{boundcond}
\end{equation}
on the plate $z=z_{1}$ and $z=z_{2}$, $z_{1}<z_{2}$, $n^{\mu }$ is
the normal to these surfaces, $n_{\mu }n^{\mu }=-1$, and $a_j$,
$b_j$ are constants. The results in the following will depend on
the ratio of these coefficients only. However, to keep the
transition to the Dirichlet and Neumann cases transparent we will
use the form (\ref{boundcond}). For the case of plane boundaries
under consideration introducing a new coordinate $y$ in accordance
with
\begin{equation}
dy=\frac{\alpha}{\eta}dz  \label{ycoord}
\end{equation}
conditions (\ref{boundcond}) take the form
\begin{equation}
\left( a_{j}+(-1)^{j-1}b_{j}\frac{\eta}{\alpha}\partial
_{z}\right) \varphi (x)=\left( a_{j}+(-1)^{j-1}b_{j}\partial
_{y}\right) \varphi (x)=0,\quad y=y_{j},\quad j=1,2.
\label{boundcond1}
\end{equation}
Note that the Dirichlet and Neumann boundary conditions are
obtained from Eq. (\ref{boundcond}) as special cases corresponding
to $(a_j,b_j)=(1,0)$ and $(a_j,b_j)=(0,1)$ respectively. The Robin
boundary condition may be interpreted as the boundary condition on
a thick plate \cite{leb}. Rewriting Eq.(8) in the following form
\begin{equation}
\varphi(x)=-(-1)^{j-1}\frac{b_{j}}{a_{j}}\partial y\varphi(x),
\end{equation}
where $\frac{a_{j}}{b_{j}}$, having the dimension of a length, may
be called skin-depth parameter. This is similar to the case of
penetration of an electromagnetic field into a real metal, where
the tangential component of the electric
field is proportional to the skin-depth parameter.\\
 Our main interest in the present paper is to investigate the vacuum
expectation values (VEV's) of the energy--momentum tensor for the field $%
\varphi (x)$ in the region $z_{1}<z<z_{2}$. The presence of
boundaries modifies the spectrum of the zero--point fluctuations
compared to the case without boundaries. This results in the shift
in the VEV's of the physical quantities, such as vacuum energy
density and stresses. This is the well known Casimir effect. It
can be shown that for a conformally coupled scalar by using field
equation (\ref{fieldeq}) the expression for the energy--momentum
tensor can be presented in the form
\begin{equation}
T_{\mu \nu }=\nabla _{\mu }\varphi \nabla _{\nu }\varphi -\frac{1}{6} \left[ \frac{%
g_{\mu \nu }}{2}\nabla _{\rho }\nabla ^{\rho }+\nabla _{\mu
}\nabla _{\nu }+R_{\mu \nu }\right] \varphi ^{2},  \label{EMT1}
\end{equation}
where $R_{\mu \nu }$ is the Ricci tensor. The quantization of a
scalar filed on background of metric Eq.(3) is standard. Let
$\{\varphi _{\alpha }(x),\varphi _{\alpha }^{\ast }(x)\}$ be a
complete set of orthonormalized positive and negative frequency
solutions to the field equation (\ref {fieldeq}), obying boundary
condition (\ref{boundcond}). By expanding the field operator over
these eigenfunctions, using the standard commutation rules and the
definition of the vacuum state for the vacuum expectation values
of the energy-momentum tensor one obtains
\begin{equation}
\langle 0|T_{\mu \nu }(x)|0\rangle =\sum_{\alpha }T_{\mu \nu }\{\varphi {%
_{\alpha },\varphi _{\alpha }^{\ast }\}},  \label{emtvev1}
\end{equation}
where $|0\rangle $ is the amplitude for the corresponding vacuum
state, and the bilinear form $T_{\mu \nu }\{{\varphi ,\psi \}}$ on
the right is determined by the classical energy-momentum tensor
(\ref{EMT1}). In the problem under consideration we have a
conformally trivial situation: conformally invariant field on
background of the conformally flat spacetime. Instead of
evaluating Eq. (\ref{emtvev1}) directly on background of the
curved metric, the vacuum expectation values can be obtained from
the corresponding flat spacetime results for a scalar field
$\bar{\varphi}$ by using the conformal properties of the problem
under consideration. Under the
conformal transformation $g_{\mu \nu }=\Omega ^{2}\eta _{\mu \nu }$ the $%
\bar{\varphi}$ field will change by the rule
\begin{equation}
\varphi (x)=\Omega ^{-1}\bar{\varphi}(x),  \label{phicontr}
\end{equation}
where for metric Eq.(3) the conformal factor is given by $\Omega
=\frac{\alpha}{\eta}$. The boundary conditions for the field
$\bar{\varphi}(x)$ we will write in form similar to Eq.
(\ref{boundcond1})
\begin{equation}
\left( \bar{a}_{j}+(-1)^{j-1}\bar{b}_{j}\partial _{z}\right) \bar{\varphi}%
=0,\quad z=z_{j},\quad j=1,2,  \label{bounconflat}
\end{equation}
with constant Robin coefficients $\bar{a}_{j}$ and $\bar{b}_{j}$.
Comparing to the boundary conditions (\ref{boundcond}) and taking
into account transformation rule (\ref{phicontr}) we obtain the
following relations between the corresponding Robin coefficients
\begin{equation}
\bar{a}_{j}=a_{j}-(-1)^{j-1}\frac{b_{j}}{\alpha},\quad
\bar{b}_{j}=b_{j}\frac{\eta}{\alpha}\label{coefrel}
\end{equation}
Note that as Dirichlet boundary conditions are conformally
invariant the Dirichlet scalar in the curved bulk corresponds to
the Dirichlet scalar in a flat spacetime. However, for the case of
Neumann scalar the flat spacetime counterpart is a Robin scalar
with $\bar{a}_j=(-1)^{j-1}\frac{1}{\eta}$ and $\bar{b}_j=1$. The
Casimir effect with boundary conditions (\ref{bounconflat}) on two
parallel plates on background of the Minkowski spacetime is
investigated in Ref. \cite{sah} for a scalar field with a general
conformal coupling parameter. In the case of a conformally coupled
scalar the corresponding regularized VEV's for the energy-momentum
tensor are uniform in the region between the plates and have the
form
\begin{equation}
\langle \bar{T}_{\nu }^{\mu }\left[ \eta _{\alpha \beta }\right] \rangle _{%
{\rm ren}}=-\frac{J_3(B_1,B_2)}{8\pi ^{3/2}a^{4}\Gamma (5/2)}{\rm
diag}(1,1,1,-3), \quad z_{1}< z< z_{2}, \label{emtvevflat}
\end{equation}
where
\begin{equation}\label{IDB1B2}
  J_3(B_1,B_2)={\rm p.v.}
\int_{0}^{\infty }\frac{t^{3}dt}{\frac{(B_{1}t-1)(B_{2}t-
1)}{(B_{1}t+1)(B_{2}t+1)}e^{2t}-1},
\end{equation}
and we use the notations
\begin{equation}
B_{j}=\frac{\bar{b}_{j}}{\bar{a}_{j}a},\quad j=1,2,\quad
a=z_{2}-z_{1}. \label{Bjcoef}
\end{equation}
For the Dirichlet scalar $B_1=B_2=0$ and one has
$J_D(0,0)=2^{-4}\Gamma (4)\zeta _R(4)$, with the Riemann zeta
function $\zeta _R(s)$. Note that in the regions $z< z_{1}$ and
$z> z_{2}$ the Casimir densities vanish \cite{{set1},{sah}}:
\begin{equation}
\langle \bar{T}_{\nu }^{\mu }\left[ \eta _{\alpha \beta }\right] \rangle _{%
{\rm ren}}=0,\quad z< z_{1},z> z_{2}.  \label{emtvevflat2}
\end{equation}
This can be also obtained directly from Eq. (\ref{emtvevflat})
taking the limits $z_{1}\rightarrow -\infty $ or $z_{2}\rightarrow
+\infty $. The values of the coefficients $B_{1}$ and $B_{2}$ for
which the denominator in the subintegrand of Eq.
(\ref{emtvevflat}) has zeros are specified in \cite{sah}. The
vacuum energy-momentum tensor on de Sitter space Eq.(3) is
obtained by the standard transformation law between conformally
related problems (see, for instance, \cite{davies}) and has the
form
\begin{equation}
\langle T_{\nu }^{\mu }\left[ g_{\alpha \beta }\right] \rangle _{{\rm ren}%
}=\langle T_{\nu }^{\mu }\left[ g_{\alpha \beta }\right] \rangle _{{\rm ren}%
}^{(0)}+\langle T_{\nu }^{\mu }\left[ g_{\alpha \beta }\right] \rangle _{%
{\rm ren}}^{(b)}.  \label{emtcurved1}
\end{equation}
Here the first term on the right is the vacuum energy--momentum
tensor for the situation without boundaries (gravitational part),
and the second one is due to the presence of boundaries. As the
quantum field is conformally coupled and the background spacetime
is conformally flat the gravitational part of the energy--momentum
tensor is completely determined by the trace anomaly and is
related to the divergent part of the corresponding effective
action by the relation \cite{davies}
\begin{equation}
\langle T_{\nu }^{\mu }\left[ g_{\alpha \beta }\right] \rangle _{{\rm ren}%
}^{(0)}=2g^{\mu \sigma }(x)\frac{\delta }{\delta g^{\nu \sigma }(x)}W_{{\rm %
div}}[g_{\alpha \beta }].  \label{gravemt}
\end{equation}
The boundary part in Eq. (\ref{emtcurved1}) is related to the
corresponding flat spacetime counterpart
(\ref{emtvevflat}),(\ref{emtvevflat2}) by the relation
\cite{davies}
\begin{equation}
\langle T_{\nu }^{\mu }\left[ g_{\alpha \beta }\right] \rangle _{{\rm ren}%
}^{(b)}=\frac{1}{\sqrt{|g|}}\langle \bar{T}_{\nu }^{\mu }\left[
\eta _{\alpha \beta }\right] \rangle _{{\rm ren}}.
\label{translaw}
\end{equation}
By taking into account Eq. (\ref{emtvevflat}) from here we obtain
\begin{equation}
\langle T_{\nu }^{\mu }\left[ g_{\alpha \beta }\right] \rangle _{{\rm ren}%
}^{(b)}=-\frac{\eta^{4}J_3(B_1,B_2)}{8\pi ^{3/2}a^{4}\Gamma (5/2)\alpha^{4}}%
{\rm diag}(1,1,1,-3),  \label{bpartemt}
\end{equation}
for $z_{1}< z< z_{2}$, and
\begin{equation}
\langle T_{\nu }^{\mu }\left[ g_{\alpha \beta }\right] \rangle _{{\rm ren}%
}^{(b)}=0,\;{\rm for}\;z< z_{1},z> z_{2}.  \label{bpartemt2}
\end{equation}
In Eq. (\ref{bpartemt}) the constants $B_{j}$ are related to the
Robin coefficients in boundary condition (\ref{boundcond}) by
formulae (\ref {Bjcoef}),(\ref{coefrel}) and are functions on
$z_j$. In particular, for Neumann boundary conditions
$B^{(N)}_j=(-1)^{j-1}\eta/a$.

The first term in Eq.(18) is the vacuum polarization due to the
gravitational field, without any boundary conditions, which
 can be rewritten as following
\begin{equation}
\langle T_{\nu }^{\mu }\left[ g_{\alpha \beta }\right] \rangle _{{\rm ren}%
}^{(0)}=-\frac{1}{2880}[\frac{1}{6} \tilde H^{(1)\mu}_{\nu}-\tilde
H^{(3)\mu}_{\nu}]
\end{equation}
 The
functions $H^{(1,3)\mu}_{\nu}$ are some combinations of curvature
tensor components (see \cite{davies}). For massless scalar field
in de Sitter space, the term is given by \cite{{davies},{Dowk}}
\begin{equation}
-\frac{1}{2880}[\frac{1}{6} \tilde H^{(1)\mu}_{\nu}-\tilde
H^{(3)\mu}_{\nu}]
=\frac{1}{960\pi^{2}\alpha^{4}}\delta^{\mu}_{\nu}.
\end{equation}
The gravitational part of the pressure according to Eq.(25) is
equal to
\begin{equation}
P_{g}=-<T^{1}_{1}>=\frac{-1}{960\pi^{2}\alpha^{4}}.
\end{equation}
This is the same from both sides of the plates, and hence leads to
zero effective force. Therefore the effective force acting on the
plates are given only by the boundary part of the vacuum pressure
, $%
p_{b}=-\langle T_{3}^{3}\left[ g_{\alpha \beta }\right] \rangle _{{\rm ren}%
}^{(b)}$, taken at the point $z=z_{j}$:
\begin{equation}
p_{bj}(z_{1},z_{2})=-\frac{\eta^{4}J_3(B_1,B_2)}{4\alpha^{4}\pi
^{3/2}a^{4}\Gamma
(3/2)}=-\frac{\eta^{4}\Lambda^{2}J_3(B_1,B_2)}{36\pi
^{3/2}a^{4}\Gamma (3/2)}. \label{vacforce}
\end{equation}
This corresponds to the attractive/repulsive force between the plates if $%
p_{bj}</>0$. The equilibrium points for the plates correspond to
the zero values of Eq. (\ref{vacforce}): $p_{bj}=0$. These points
are zeros of the function $J_3(B_1,B_2)$ defined by Eq.
(\ref{IDB1B2}) and are the same for both plates.

\section{Parallel plates with different cosmological constants
 between and out-side}

 Now, assume there are different vacuua between and out-side of
 the plates, corresponding to $\alpha_{betw}$ and $\alpha_{out}$ in the
 metric Eq.(3). As we have seen in the last section, the vacuum
 pressure due to the boundary is only non-vanishing between the
 plates. Therefore, we have for the pressure due to the boundary
 \begin{equation}
 p_{bj}(z_{1},z_{2})=-\frac{\eta^{4}J_3(B_1,B_2)}{4\alpha_{betw}^{4}\pi
^{3/2}a^{4}\Gamma (3/2)}=-\frac{\eta^{4}\Lambda_{betw
}^{2}J_3(B_1,B_2)}{36\pi ^{3/2}a^{4}\Gamma (3/2)}.
 \end{equation}
 Now, the effective pressure created by gravitational part Eq.(25), is
 different for different part of the space-time:
 \begin{equation}
 P^{betw}_{g}=-<T^{1}_{1}>_{betw}=\frac{-1}{960\pi^{2}\alpha_{betw}^{4}}=
 \frac{-\Lambda_{betw}^{2}}{9}\frac{1}{960\pi^{2}},
 \end{equation}
 \begin{equation}
 P^{out}_{g}=-<T^{1}_{1}>_{out}=\frac{-1}{960\pi^{2}\alpha_{out}^{4}}=
 \frac{-\Lambda_{out}^{2}}{9}\frac{1}{960\pi^{2}}.
 \end{equation}
 Therefore, the gravitational pressure acting on the plates is
 given by
 \begin{equation}
 P_{g}=P^{betw}_{g}-P^{out}_{g}=\frac{-1}{9\times 960\pi^{2}}
 (\Lambda_{betw}^{2}-\Lambda_{out}^{2}).
 \end{equation}
 The total pressure acting on the plates, $P$, is then given by
 \begin{equation}
 P=P_{g}+P_{b}=\frac{-1}{9\times 960\pi^{2}}
 (\Lambda_{betw}^{2}-\Lambda_{out}^{2})-\frac{\eta^{4}\Lambda_{betw
}^{2}J_3(B_1,B_2)}{36\pi ^{3/2}a^{4}\Gamma (3/2)}.
 \end{equation}
 In figures $1, 2$, I have plotted the vacuum boundary pressure (second summand
in Eq.(32)) acting per unit surface of the plates as functions of
$a_p/\tilde {B_2}$ where $a_p=a \alpha_{betw} /\eta$ is the proper
distance between the plates, which is given by
\begin{equation}
a_{p}=\int_{z_{1}}^{z_{2}}\sqrt{-g_{33}}dz=a \alpha_{betw} /\eta
\end{equation}
and, $\tilde {B_i}=B_i/a_p$ (note that $\tilde {B_i}$ do not
depend on $\eta$ and $a$). Fig1  for $\tilde {B_1}=0$ and fig2 for
$\tilde {B_1}=\tilde {B_2}$. The first case is also presented in
paper by Saharian and Romeo \cite{sah}.

As a result we have an example for the stabilization of the
distance between the plates due to the vacuum boundary pressures.
 But in this case total pressure may be negative or positive. To see the different
  possible cases, let us first assume a false
 vacuum between the plates, and true vacuum out-side, i.e. $\Lambda_{betw}>
 \Lambda_{out}$, then the gravitational part is negative,
 $p_{g}<0$, depending to the boundary part pressure the following
 cases can be occur
\begin{equation}
P_{b}>0, \hspace{0.5cm} P_{b}>|p_{g}|  \Rightarrow p>0,
\end{equation}
\begin{equation}
P_{b}>0, \hspace{0.5cm} P_{b}=|p_{g}|  \Rightarrow p=0,
\end{equation}
\begin{equation}
P_{b}>0, \hspace{0.5cm} P_{b}<|p_{g}|  \Rightarrow p<0,
\end{equation}
\begin{equation}
P_{b}=0  \Rightarrow p<0, \hspace{1cm}P_{b}<0  \Rightarrow p<0.
\end{equation}
  For the case of true vacuum between
 the plates and false vacuum out-side, i.e.
 $\Lambda_{betw}<\Lambda_{out}$, the gravitational pressure is
 positive $p_{g}>0$. In this case the following possibility can be
 occur
\begin{equation}
P_{b}<0, \hspace{0.5cm}|P_{b}|<p_{g}  \Rightarrow p>0,
\end{equation}
\begin{equation}
P_{b}<0, \hspace{0.5cm}|P_{b}|>p_{g}  \Rightarrow p<0,
\end{equation}
\begin{equation}
P_{b}<0, \hspace{0.5cm}|P_{b}|=p_{g}  \Rightarrow p=0,
\end{equation}
\begin{equation}
P_{b}>0,   \Rightarrow p>0,\hspace{1cm}P_{b}=0 \Rightarrow p>0.
\end{equation}
As one can see in Eqs.(35, 40) the boundary part and gravitational
part pressure cancel each other out, in this case the plates are
fixed and we have stable equilibrium points. In the cases with
$p</>0$, the initial attraction /repulsion of the parallel plates
may be stopped or not depending on the detail of the dynamics.
\begin{figure}[tbph]
\begin{center}
\epsfig{figure=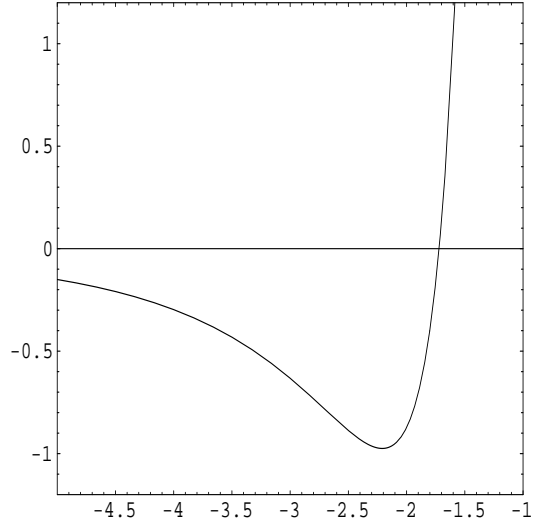,width=7cm,height=7cm}
\end{center}
\caption{Boundary part of pressure  as a function on
$a_p/\tilde{B_2}$ , when $\tilde{B_1}=0$.} \label{fig1ID}
\end{figure}

\begin{figure}[tbph]
\begin{center}
\epsfig{figure=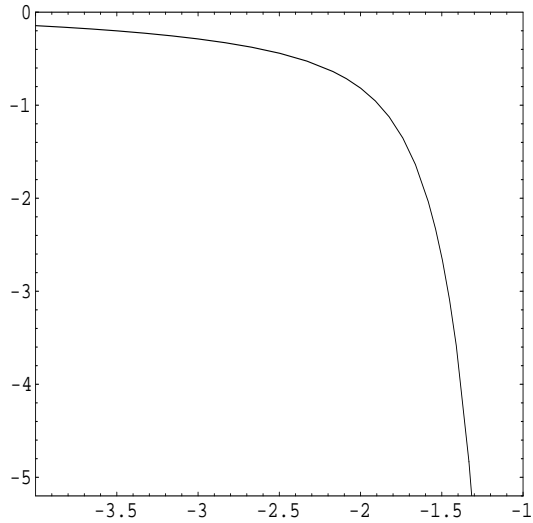,width=7cm,height=7cm}
\end{center}
\caption{Boundary part of pressure  as a function on
$a_p/\tilde{B_2}$ , when $\tilde{B_1}=\tilde{B_2}$. }
\label{fig2zer}
\end{figure}

 \section{Conclusion}
In the present paper we have investigated the Casimir effect for a
conformally coupled massless scalar field confined in the region
between two parallel plate with constant comoving distance on
background of the conformally-flat de Sitter spacetimes. The
general case of the mixed(Robin) boundary conditions is
considered. The vacuum expectation values of the energy-momentum
tensor are derived from the corresponding flat spacetime results
by using the conformal properties of the problem. In the region
between the plates the boundary induced part for the vacuum
energy-momentum tensor is given by Eq.(22), and the corresponding
vacuum forces acting per unit surface of the plates have the form
Eq. (27). These forces vanish at the zeros of the function
$J_3(B_1,B_2)$.The vacuum polarization due to the gravitational
field, without any boundary conditions is given by Eq.(25), the
corresponding gravitational pressure part has the form Eq.(26),
which is the same from both sides of the plates, and hence leads
to zero effective force. Further we consider different
cosmological constants for the space between and outside of the
plates, in this case the effective pressure created by
gravitational part is
 different for different part of the space-time and add to the
 boundary part pressure.
  Our calculation shows that the detail dynamics of the plates
  depends on different parameters and all cases of attraction and repulsion
  may appear.  The result may be of interest in the case of
 formation of the cosmic domain walls in early universe, where the
  wall orthogonal to the $z-$axis is described by the
 function $\Phi_{i}(z)$ interpolating between two different minima
 at $z\rightarrow \pm\infty$ \cite{vil1}.

\section*{Acknowledgement }
I would like to thank Prof. Saharian for his help in plot of
graphs.

\end{document}